# Document watermarking based on digital holographic principle

Chol-Su Kim, Kwang-Hyok Jong, Song-Jin Im

Institute of Optics, Department of Physics, **Kim Il Sung** University, Pyongyang, DPR of Korea

**Abstract;**

A new method for document watermarking based on the digital Fourier hologram is proposed. It applies the methods of digital image watermarking based on holographic principle presented previously in several papers into printed documents. Experimental results show that the proposed method can not only meet the demand on invisibility, robustness and non-reproducibility of the document watermark, and but also has other advantages compared with the conventional methods for document securities such as embossed hologram, Lippmann photograph and halftone modulation.

1. Introduction

The demand on document watermarking has been raised since long ago for security, anticounterfeiting and antiforgery. Its typical example is the watermarks in banknotes. There are also needs for watermarking for personalized documents, such as passports, identification cards, travel documents, driving licenses, credit cards, etc. On the other hand, some institutions and companies do not want their documents to be illegally distributed without any permission and/or want to claim the ownership for their documents. For this reason, they need document watermarking in visible or invisible way. There might be many kinds of physical and chemical watermarking methods. But, our considerations are restricted only to the optical one.

Up to now mass-produced embossed holograms or other types of mass-produced Optical variable devices (OVDs) are widely used in the field of document security [1]. The recent introduction of color holography offers a possibility to apply full color volume reflection holograms in the field of document security. Another technique, interferential photography of Lippmann photography, represents a new type of OVD, which belongs to the interference security image structures. These techniques require sophisticated equipment and special recording materials to produce OVDs. Thus their applications are limited to certain documents.

Other possibilities for document security are to use the optical information processing. The typical one of them is to encode visual information in a halftone image using the locations of the dots inside their cells [2]. The algorithm puts together two data files for two images. One is the observable picture to print and the other one is the hidden image within the observable picture. Once an image is encoded, only an authorized person who has the key can reveal the hidden image. The composite image can be printed on any printer. The print can then be read by a conventional optical scanner and processed by computer, or optical correlator, to access the hidden image. In general, the quality of halftone image is much lower than grey scale image. Therefore, it is not to be avoidable that the quality of the printed document of halftone image is getting worse than corresponding grey scale image.

In this paper, a new method for document watermarking based on digital holographic principle is proposed. Up to now several watermarking methods by use of digital holography have been known [4, 5, 6]. The hidden image is embedded into the content image to be protected in a form of diffused digital hologram. Every part of the visible content image contains information on the entire hidden image, such that if one covers or destroys part of the image the entire hidden image can still be recovered from the rest. But, their applications have still been restricted within digital images. The image with watermark can be printed on any printer. The printed document can then be read by a conventional optical scanner or digital image captures such as digital cameras and processed holographically to extract the hidden image. It has the same properties as the digital holographic watermarking does. Furthermore, there are several another key points to decode the watermarked document such as original size in pixel, resolution in dpi and positioning. Without knowing them, nobody can reconstruct the watermark image correctly.

## 2. Holography for digital image watermarking

In the watermarking method proposed by Takai and Mifun [4], the image to be hidden has to be converted into a digital hologram before the embedding process. The digital construction and its reconstruction are described in this section.

**A. Simulation of digital hologram**

First of all, the mark image to be hidden is resized and inserted into a zero-image with the same dimensions as the content image to be protected. The Fourier hologram is stimulated with this zero-padded image.

If an object with diffuse surface is illuminated by collimated light wave or even by random light wave, the light wave is scattered into all the possible directions. This scattering character of the reflected (or transmitted) light wave can be simulated with random phase mask.

$$g_o(x, y) = g_m(x, y)\exp[i\phi(x, y)] \tag{1}$$

where $g_m(x, y)$ is the zero-padding image with mark image and $\phi(x, y)$ is 2-D random number within the range [0,1]. The diffuse Fourier hologram $H_o(\xi,\eta)$ of the mark image is formed with interference between the Fourier transformation of Eq. (1) $G_m(\xi,\eta)$ and planar reference wave $R(\xi,\eta)$.

$$G_m(\xi,\eta) = \int_{-\infty}^{\infty}\int_{-\infty}^{\infty} g_o(x,y)\exp[-2\pi i(a\xi+b\eta)]dxdy \tag{2}$$

$$R(\xi,\eta) = R_o \exp[2\pi i(a\xi+b\eta)] \tag{3}$$

The parameters *a* and *b* in Eq. (3) are the angular spectrums of the planar reference wave.

$$\begin{aligned}H_o(\xi,\eta) &= |G_m(\xi,\eta)+R(\xi,\eta)|^2 \\ &= |G_m(\xi,\eta)|^2 + |R(\xi,\eta)|^2 \\ &\quad + G_m^*(\xi,\eta)R(\xi,\eta) + G_m(\xi,\eta)R^*(\xi,\eta),\end{aligned} \tag{4}$$

where the symbol $^*$ denotes the complex conjugate. The two first terms in left side of Eq.(4) are neglected because they don't carry any useful information.

$$H(\xi,\eta) = G_m^*(\xi,\eta)R(\xi,\eta) + G_m(\xi,\eta)R^*(\xi,\eta) \tag{5}$$

$H(\xi,\eta)$ in Eq. (5) is the digital hologram which will be superposed onto the content image.

If Eq.(5) is inversely Fourier-transformed and let $R_o$ in Eq.(3) be 1, two images are reconstructed, which are symmetrical and conjugated to each other.

$$g_r(\xi,\eta) = \int_{-\infty}^{\infty}\int_{-\infty}^{\infty} H_o(\xi,\eta)\exp[2\pi i(a\xi+b\eta)]d\xi d\eta \tag{6}$$

$$g_r(x,y) = g_o^*(x-a, y-b) + g_o[-(x+a),-(y+b)] \tag{7}$$

The watermark image is real, so the intensity signal obtained in the reconstructed plane is expressed as

$$\begin{aligned}|g_r(x,y)|^2 &= |g_o^*(x-a, y-b)|^2 + |g_o[-(x+a),-(y+b)]|^2 \\ &= |g_m^*(x-a, y-b)|^2 + |g_m[-(x+a),-(y+b)]|^2\end{aligned} \tag{8}$$

The positions of the two watermark images (*a, b*) and (-*a,-b*) in the reconstruction plane can be controlled by selecting the position of mark image in zero padding and controlling the angular spectrums *a* and *b* of reference wave during recording process.

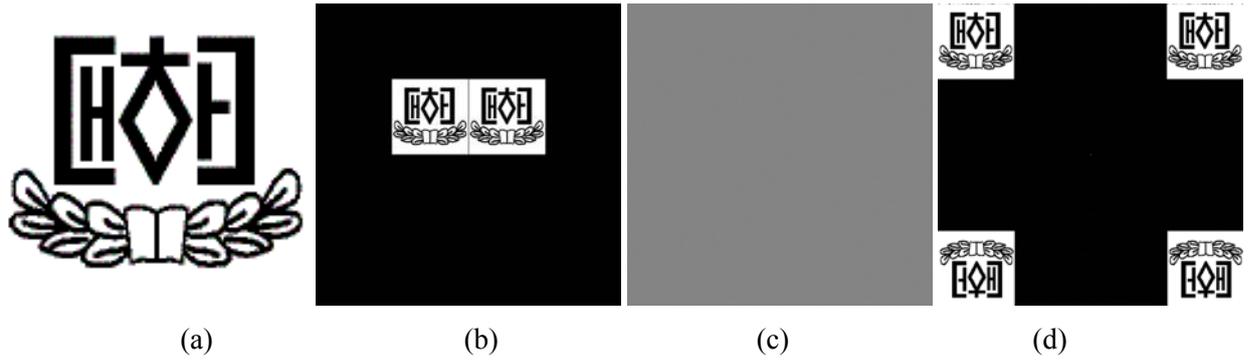

(a)　　　　　　　　　(b)　　　　　　　　　(c)　　　　　　　　　(d)

Fig.1. (a) Mark image to be hidden (128 × 128 pixels), (b) zero-padding of mark image (512 × 512 pixels), (c) Fourier hologram of (b), (d) reconstructed image.

Figure 1 shows the mark image to be hidden which is the cap badge of **Kim Il Sung** University, the mark image within zero-padding, Fourier hologram and its reconstruction. The mark images are resized to a sixteenth of the dimension of content image and laid side by side.

**B. Low-pass-filtering of content image**

Usually, the most intensity of Fourier spectra of digital images is concentrated in low frequency region. That is the central region of Fourier frequency domain. It means that the high frequency regions can be used for reconstruction of hologram and the content image accordingly has to be low-pass filtered. The high frequency regions at the 4 corners of Fourier spectrum of content image are set to be zero during low-pass filtering. In this way all high frequency information is eliminated from the image. This operation is necessary because the mark can be extracted from the watermarked image spectra. For this reason the spectra of the filtered content image and the spectra of the digital hologram have to be spatially separated. By means of the filtering, the content image spectrum is concentrated only in the low and medium frequencies, whereas the hologram spectrum is only in the high frequency. It is important to prevent the overlap effect between the watermark image and the spectrum of the content image. The size of a high frequency region is a sixteenth of one of the content image. The location of the reconstructed images can be controlled by setting the parameters *a* and *b* defined in Eq. (3). For example, the size of a high frequency region is 128 × 128 pixels for the content image with dimension of 512 × 512 pixels.

**C. Watermarking process based on digital hologram**

In the watermarking process the hologram is simply superposed onto the low-pass filtered content image with a certain weight. Before superposing, both the pixel values in the digital hologram and content image have to be normalized into the range [0, 1] to maintain the content image quality as good as possible. This method can be applied onto a gray scale image or to a RGB true color image, operating on each color channel.

Figure 2 shows an example of digital watermarking based on holographic principle and its reconstruction, where the weight factor of hologram was 0.1.

**3. Document watermarking using digital holography**

The applications of Takai and Mifun's method have still been restricted within digital images [4, 5, 6]. The digital image with watermark can be printed with any printer. The printed document can then be read by a conventional optical scanner or digital image captures such as digital cameras and processed holographically to extract the hidden image.

Figure 3 shows the schematic diagram of the proposed method for document watermarking and reconstruction. The document to be protected is holographically watermarked in computer by Takai and Mifun's method and then printed. This document has to be re-digitized by image capture devices like scanner or digital camera to extract the hidden mark image. There are several key points to decode the watermarked document correctly such as original image size in pixel, printing resolution in dpi and its position.

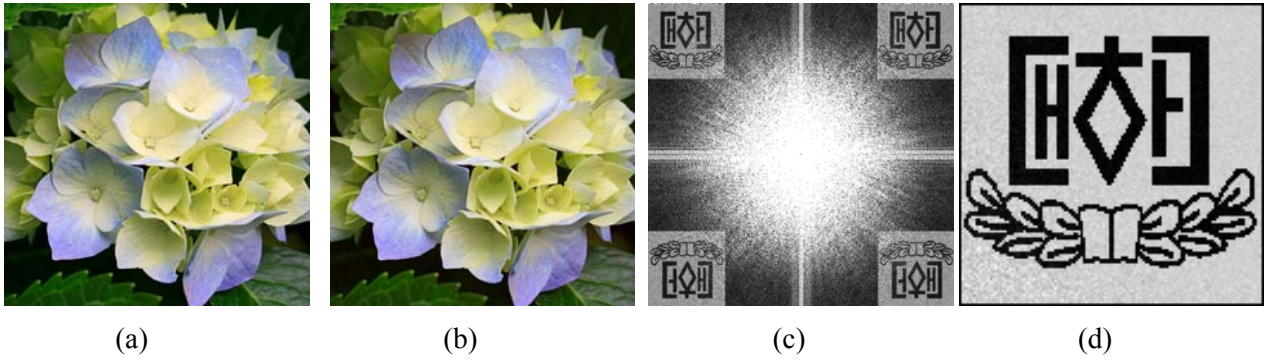

(a)　　　　　　　　　(b)　　　　　　　　　(c)　　　　　　　　　(d)

Fig.2. (a) content image (512 × 512 pixels), (b) watermarked low-pass filtered image(weight factor of hologram was 0.1), (c) its Fourier spectrum, (d) extracted mark image((128 × 128 pixels)

Without knowing them, nobody can reconstruct the correct watermark image. For example, if the size of re-digitized image differs than one of the original image, the mark image cannot be reconstructed correctly.

Watermarking with digital holographic techniques has the problem that the mark can be easily recovered with a Fourier transform and then the content image can be tampered with and be watermarked with the same holographic pattern. To avoid this situation the pixel scrambling with private key encryption scheme introduced by Spagnoloa *et al* can be used to enhance the security level of the watermarked images [6].

**4. Experimental results**

In our experiments, RGB true color content images with dimension 512 × 512 pixels have been used. The images have been low-pass filtered to separate mark images from the Fourier spectrum of content images. The mark images were black and white with dimension equal to 128 × 128 pixels (see figure 1). Actually, there is no limitation for content image and mark image dimensions; in fact the mark images have to be simply resized to a quarter of the width and a quarter of the height of the content image. The printer and scanner used in our experiments were normal commercial laser printer and scanner.

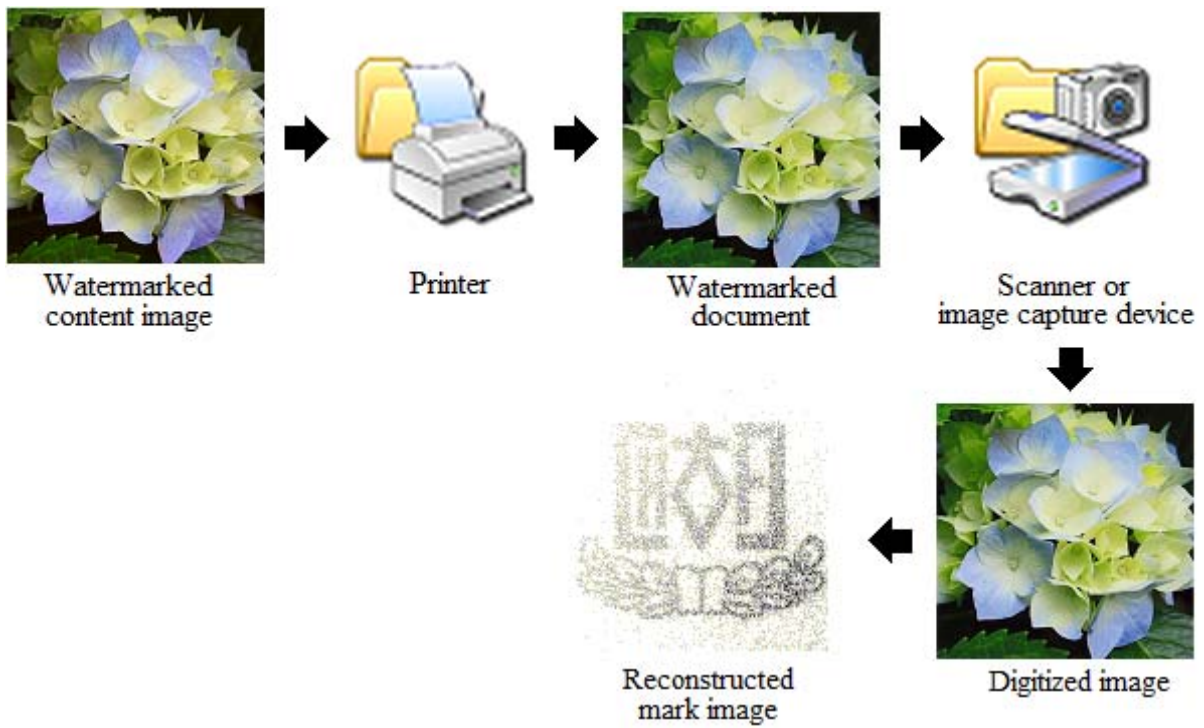

Fig.3. Schematic diagram of document watermarking and watermark extraction

The sizes of documents can be set with printing resolution in dpi unit. For example, a content image with dimension 512 × 512 pixels is printed to the document with size 13.5cm ×13.5cm in printing resolution 96dpi.

Figure 4 shows the digital watermarked content image, the re-digitized image of the document printed from it and reconstructed mark, where the weight constant for hologram during digital watermarking was 0.2. Comparing Figure 4(c) with Figure 2(d), the mark image reconstructed from the document is worse than that one from the digital image. But, it is still clearly recognizable.

In digital holographic watermarking, it is possible to reconstruct the mark image not only from full images but also from any partial one (see figure 5). It means that we can verify our document even from its any optional part.

Because of the property of Fourier hologram [3], the position of partial images in the field of content image doesn't affect the reconstructed mark image (see figure 6). It is also very important in verifying the worth of documents because we don't need to be careful of the original positions of partial images in the field of content image.

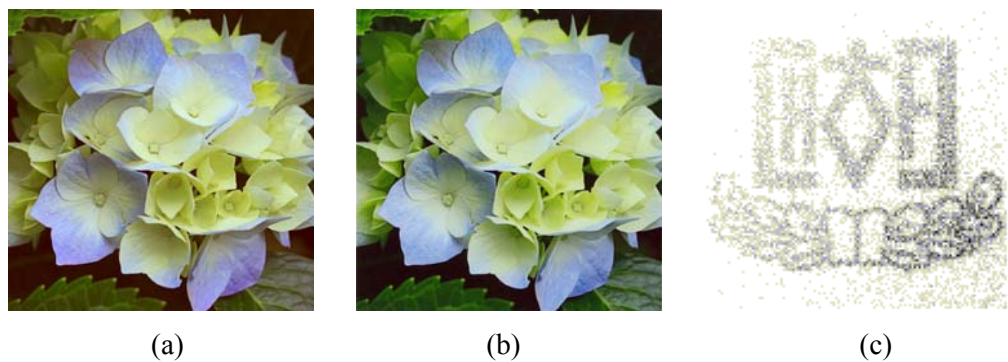

(a)　　　　　　　　(b)　　　　　　　　(c)

Fig.4. (a) Digital watermarked content image, (b) The re-digitized image of the document printed from (a), (c) Reconstructed mark

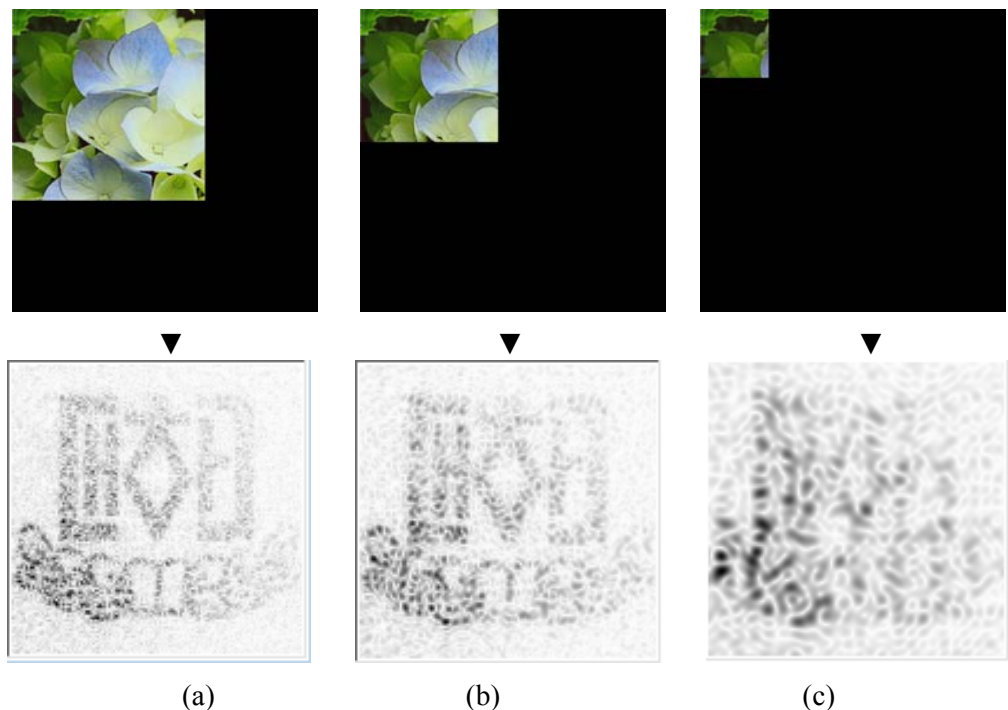

(a)　　　　　　　　(b)　　　　　　　　(c)

Fig.5. Partial images and corresponding reconstructed marks. (a) 40%, (b) 20%, (c) 5%

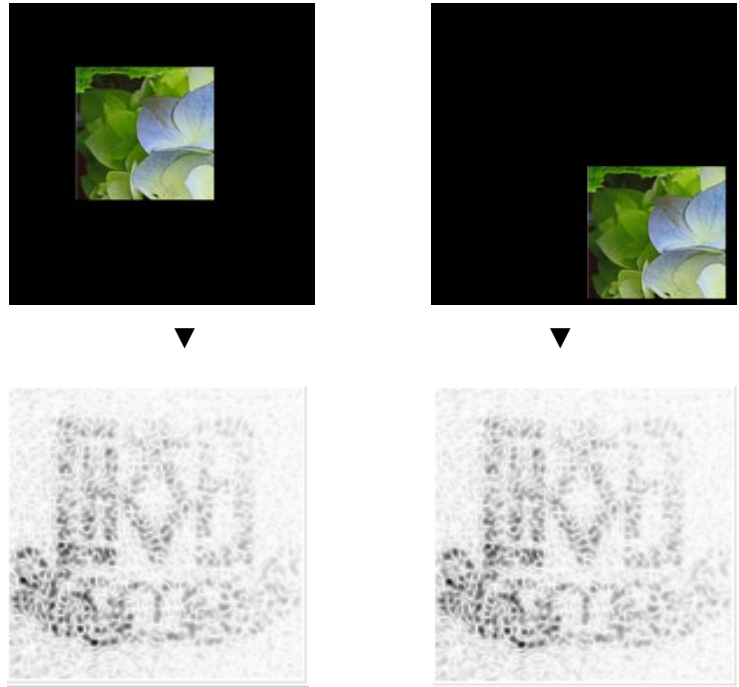

Fig.6. Partial images in different positions and corresponding reconstructed marks

There could be angular difference between content watermarked image before printing and re-digitized image. The mark image is still retrievable in presence of angular difference of a few degrees. But, when pixel scrambling is introduced as mentioned above, no angular difference is allowed to retrieve the mark correctly. It means that only the person who embedded the mark image can reconstruct it because he knows the position of document before printing and how to position it in scanning. It is useful to confirm the security of document watermarking.

**5. Conclusion**

We have proposed an application ability of digital holographic watermarking into document security. The digital holographic watermarking is based on Takai and Mifun's method and can also be applied onto a gray scale image or to a RGB true color image, operating on each color channel. The holographically watermarked content image is printed with any printer. The holographic pattern embedded into content image which is controllable with weight factor during watermarking process is transmitted onto printed document. The document can then be read by a conventional optical scanner or digital image captures such as digital cameras and processed holographically to extract the hidden image. The reconstructed mark image is not so good as one from the original digital content image because there have been some information losses during printing and scanning. But, it is still clear enough to recognize.

There are several key points to encode the watermarked content image such as original size in pixel, resolution in dpi and position of document. Without knowing them, nobody can reconstruct the correct watermark image. To improve of the watermarking security, pixel scrambling with random number generator can be used.

This method is cheap, economic, flexible and effective, and can be used successfully where the security of document is necessary.

**References**

[1] H. I. Bjelkhagen, "New OVDs for Personalized Documents Based on Color Holography and Lippmann Photography," Optical and Digital Techniques for Information Security, ed. by B. Javidi, Springer Science+Business Media, Inc. , 2005, pp. 17-36

[2] J. Rosen and B. Javidi, "Steganography and Encryption Systems Based on Spatial Correlators with Meaningful Output Images" Optical and Digital Techniques for Information Security, ed. by B. Javidi, Springer Science+Business Media, Inc. , 2005, pp. 59-94
[3] J. R. Collier, C. B. Burckhardt and L. H. Lin, "Optical Holography" , Academic Press,1971, Chap. 8, pp. 209
[4] Takai N and Mifune Y , "Digital watermarking by a holographic technique," Appl. Opt. 41, 865–73(2002)
[5] Hsuan T. Chang and Chung L. Tsan, "Image watermarking by use of digital holography embedded in the discrete-cosine-transform domain" Appl. Opt. 44, 6211–19(2005)
[6] G. S. Spagnolo, C. Simonetti and L. Cozzella, "Content fragile watermarking based on a computer generated hologram coding technique," J. Opt. A: Pure Appl. Opt. **7,** 333–42(2005)